\documentclass[prd,aps,superscriptaddress,floatfix,nofootinbib,eqsecnum,notitlepage]{revtex4-1}
\pdfoutput=1

\usepackage{amsfonts}
\usepackage{amsmath}
\usepackage{amssymb}
\usepackage{tikz}
\usepackage{bm}
\usepackage{dcolumn}
\usepackage{graphicx}   
\usepackage[utf8]{inputenc}
\usepackage{latexsym}
\usepackage{rotating}
\usepackage{hyperref}
\usepackage{subfigure}
\usepackage{color}
\usepackage{changes}

\begin{document}

\title{Equation of state to compact objects with ion crystal lattice}

\author{S.~V.~C.~Ramalho}
\email{saulovcr@gmail.com}
\affiliation{CBPF - Centro Brasileiro de
Pesquisas F\'{\i}sicas, Xavier Sigaud st. 150,
zip 22290-180, Rio de Janeiro, Brazil.}

\author{S.~B.~Duarte}
\email{sbd.cbpf.rj@gmail.com}
\affiliation{CBPF - Centro Brasileiro de
Pesquisas F\'{\i}sicas, Xavier Sigaud st. 150,
zip 22290-180, Rio de Janeiro, Brazil.}

\author{G.~S.~Vicente}
\email{gustavo@fat.uerj.br}
\affiliation{FAT - Faculdade de Tecnologia, UERJ - Universidade do Estado do Rio de Janeiro, Pres. Dutra rd. km 298,
	zip 27537-000, Resende, Brazil.}

\begin{abstract}
Crystal lattice structure is present in stellar compact objects, such as white dwarf stars and in the crust of neutron stars.
These structures can be described by a body-centered cubic crystal, which is formed by ions due to Coulombian interactions in presence of stellar electron plasma. 
The electron-electron interaction is currently described in the electrodynamics context by a quantum homogeneous plasma. Particularly, we investigate the changes in the medium Equations of State (EoS), improving results presented in the literature.
The main purpose of this work is to analyse the distribution of electrons in the stellar medium considering their interaction with  ions in the  crystal lattice. 
The electric field produced by the presence of crystal lattice is obtained using linear response theory in the context of finite temperature field theory. 
The screening and distribution of electrons are corrected by an arbitrary number of neighbor ions, which is a novelty in the literature and significantly impact the EoS. 
Numerical results are presented for a completely degenerate electron plasma and for different species of ions that make up the lattice. 
These EoS can be applied to determine the structure of the aforementioned compact stellar objects.
\end{abstract}

\maketitle 

\section{Introduction}
\label{sec1}

At the end of the evolution cycle of a star, high energy phenomena take place and it eventually becomes a supernovae or a red giant, from which it turns a compact object. 
This object can be a black hole, a white dwarf or a neutron star, and is characterized by its great density~\cite{Glendenning,Camenzind}. 
White dwarfs and neutron stars are massive objects with different masses, sizes and density scales. White dwarfs are approximately Earth-sized and have mass densities that reach $\rho \sim 10^6\, {\rm g/cm}^{3}$, while neutron stars have radius around $R\sim 10\, {\rm km}$ and densities that reach supranuclear density scales.
These objects reach extreme stellar conditions of temperature, density and pressure as well as strong magnetic fields and fast rotations. 
Compact objects are of great interest as they can be seen as excellent laboratories for observations of new phenomena involving strong magnetic fields as well as high densities and temperatures, which cannot be reproduced in Earth-based experiments.

In order to understand the formation and structure of these objects, it is necessary to build the stellar matter Equation of State (EoS). 
There are many approaches in the literature dedicated to obtain the EoS for such objects~\cite{Haensel:2007yy,Chamel:2008ca, Lattimer:2000nx, measuriment, Ozel:2016oaf}.
However, even with such progress, this subject is still not completely understood.

Particularly, neutron star outer crusts and white dwarfs exhibit two similar aspects in their structures.
Firstly, there are regions composed by a plasma of electrons and positrons, which we will refer to throughout this work simply as plasma.
Secondly, both objects present a crystal lattice in its structures~\cite{Haensel:2007yy,Chamel:2008ca, baiko}. 
In the case of white dwarfs, the presence of a crystal lattice in its cooling core is supported by recent observational evidences~\cite{nature}.

The presence of a crystal lattice in these objects induces a rearrangement of charges in the plasma, with  
ions in the lattice screened by the electron cloud.
As a consequence, the equilibrium configuration is achieved at uniform temperature, leading to a position dependent chemical potential of the plasma.
Therefore, the crystal lattice can impose relevant changes in the EoS of the plasma. 

The quantum field theory approach for the EoS of the plasma in the presence of a lattice of ions is carried out concerning electrodynamics interactions~\cite{bellac,kapusta}. The screening of the plasma is usually described by simplified semi-classical approaches, such as the Feynman-Metropolis-Teller (FMT) one~\cite{Rotondo:2011zz}. 
The screening and electron distribution are here corrected by including the ionic lattice structure and contributions of ions in high level of ion neighbourhood to the electron chemical potential.  
The calculations are done using quantum field theory at finite temperature~\cite{kapusta,bellac}, where the ion's interactions are treated as external classical perturbations by means of linear response theory. The crystal cell is considered as electrically neutral, although the stable non homogeneous electron distribution inside the cell induces a high order multi-polarity potential in the neighborhood.

We remark that the study of neutrino emission from the surface of neutron star outer crust by plasmon decay is a very relevant aspect of the neutron star cooling~\cite{Baldo:2009we,Braaten:1993jw,Yakovlev:2000jp,Altherr:1993hb}. 
Consequently, the improvement of the EoS including the ion lattice interaction with the plasma may affect the star cooling process.  

The paper is organized as follows.
In Sec.~\ref{sec2}, we report the main aspects of the usual EoS for a QED plasma in a homogeneous medium.
The EoS are presented including exchange interactions and for different limits for mass and temperature. 
In Sec.~\ref{sec3},
we briefly introduce the crystal lattice structure and the one-component plasma (OCP) and multi-component plasma (MCP)
models.
We also present the Coulomb coupling parameter, that is used in literature to describe the condition of the crystal lattice formation.
In Sec.~\ref{sec4},
we introduce the screened electric field generated by the lattice ions when immersed in the QED plasma.
The electric field of lattice ions is introduced as an external field applied to the plasma by means of linear response theory.
We write the ion density for a body-centered cubic (bcc) crystal in the MCP model and consider in our study the OCP model as a particular case.
The electrical potential generated by the lattice ions is interpreted as a correction to the electron chemical potential, resulting in an electrochemical potential for the plasma. 
In Sec.~\ref{sec5}, we discuss a degenerate plasma gas in both non-relativistic and relativistic limits
and provide the EoS inside the Wigner-Seitz (WS) cell for each of these scenarios. 
In Sec.~\ref{sec6}, 
we introduce macro thermodynamics variables for each cell, which are important to determine the EoS of the plasma for future applications in compact objects. 
These variables provide the effective behavior of the position dependent variables of Sec.~\ref{sec5} in larger scales of the medium.
We also define the WS cell in its exact polyhedral description, introduce the spherical approximation commonly used in the literature and calculate the WS radius for arbitrary number of neighbor ions, which is discussed in details in
the Appendix~\ref{appendixrcellsec}.
In Sec.~\ref{result}, we present representative numerical results which illustrates the non homogeneous distribution of chemical potential, number density and pressure of the plasma in the interior of primitive cells and EoS for the macro thermodynamic variables 
for the relativistic degenerate gas limit. 
We also present a comparison of our results with the ones of other models in the literature, such as the models introduced by Chandrasekhar (Ch)~\cite{Chandrasekhar:1931ih},
Salpeter (S)~\cite{Salpeter:1961zz} and Feynman-Metropolis-Teller~\cite{Rotondo:2011zz}.
We address the electron capture and neutron drip limit, which are important aspects for the applicability of our model.
Finally, in Sec.~\ref{sec7} we draw some general conclusions about our results and perspectives of applications. 
Details of calculations are left to appendices. 
In Appendix~\ref{equilibriumappendix}, we present the new local hydrostatic equilibrium achieved after the introduction of the crystal lattice in the electron gas.

\section{Homogeneous QED plasma with electron-electron exchange interactions}
\label{sec2}

Here we describe the electron-positron plasma in the framework of QED at finite temperature. 
The QED exchange interactions is represented by the temperature dependent functional $\mathcal{Z}$, 
from which the EoS for the homogeneous plasma is derived. 
However, due to the difficulty in obtaining analytical results, we have to consider some limits and derive approximate expressions.  
The preliminary step is to introduce $\ln \mathcal{Z}_{0}$ for QED, where $\mathcal{Z}_{0}$ is the free generating functional (no exchange interactions) and is given by~\cite{kapusta}: 
\begin{eqnarray}\label{Z0}
\ln \mathcal{Z}_{0}
=
2V
\int 
\frac{d^3 k}{(2\pi)^3}
&&\left\{ 
\ln\left[
1+ e^{-\beta(E-\mu_e)}\right]
+
\ln\left[
1+ e^{-\beta(E+\mu_e)}\right]
\right\},
\end{eqnarray}
where the first (second) logarithm refers to electrons (positrons), 
$V$ is the volume,
$\beta=1/T_e$, 
$T_e$ is the temperature,
$\mu_e$ is the chemical potential, $E_e=\sqrt{p_e^2+m_e^2}$ is the energy, $p_e$ is the momentum and $m_e$ is the mass~\footnote{The subscript ``e" in the physical quantities refers to the electron.}.
In terms of the generating functional, the  pressure EoS is defined by:
\begin{eqnarray}\label{P}
P=\frac{\ln \mathcal{Z}}{\beta V}.
\end{eqnarray}
In order to write analytical results, we consider some limiting cases with respect to electron mass $m_e$ and temperature $T_e$. 
These limits are the degenerate ($T_e\ll \mu_e-m_e$) and non-degenerate gases ($T_e\gg \mu_e-m_e$).
In the former limit, we also consider the non-relativistic ($p_F\ll m_e$) and the relativistic ($p_F\gg m_e$) cases, and in the latter we consider the ultrarelativistic case ($T_e,\mu_e\gg m_e$).

In the free case ($\mathcal{Z}=\mathcal{Z}_0$), 
from Eq.~\eqref{P}, in the degenerate limit ($T_e=0$) the pressure results:
\begin{eqnarray}\label{p0deg}
P_0=\frac{1}{16\pi^2}\left[p_F\mu_e(2\mu_e^2-m^2_e )-m^4_e\ln\left(\frac{\mu_e+p_F}{m_e}\right)\right], 
\end{eqnarray}
where $p_F=\sqrt{\mu_e^2-m_e^2}$ is the Fermi momentum.
On the other hand, for a the non-degenerate gas in the ultrarrelativistic limit ($m_e=0$), the pressure results:
\begin{eqnarray}\label{p0t}
P_0=\frac{\mu_e^4}{12\pi^2}+\frac{\mu_e^2T_e^2}{6}+\frac{7\pi^2T_e^4}{180} .
\end{eqnarray}

In order to include the exchange interactions, one considers the interaction term $\ln \mathcal{Z}_{I}$, given by Eqs.~(5.59) and~(5.60) of Ref.~\cite{kapusta}.
Therefore, we now write $\ln\mathcal{Z}=\ln\mathcal{Z}_0+\ln\mathcal{Z}_I$ in Eq.~\eqref{P}.
We also define the pressure $P_{e-e}=\ln \mathcal{Z}_{I}/(\beta V)$, which is the pressure contribution due to exchange interactions.
In the degenerate limit, one obtains:
\begin{eqnarray}\label{pideg} 
P_{e-e}=
-\frac{e^2}{(2\pi)^4}
\left\{\frac{3}{2}\left[\mu_e p_F-m_e^2\ln\left(\frac{\mu_e+p_F}{m_e}\right)\right]^2-p_F^4\right\},
\end{eqnarray}
whereas in the non-degenerate  ultrarrelativistic limit one obtains:
\begin{eqnarray}\label{pit}
P_{e-e}=-\frac{e^2}{288}\left(5T_e^4+\frac{18\mu_e^2T_e^2}{\pi^2}+\frac{9\mu_e^4}{\pi^4}\right). 
\end{eqnarray}
Writing the total pressure as $P_e=P_0+P_{e-e}$, we obtain the EoS for the pressure including exchange interactions~\footnote{The authors of Ref.~\cite{Hossain:2019teg} present a slightly different result from the one given in Ref.~\cite{kapusta} for Eq.~\eqref{pideg}, which produces a difference of around $0.7\%$ in the results for the Wigner-Seitz radius. We have considered the latter result in this paper.},
which in the degenerate limit it reads:
\begin{eqnarray}\label{pt}
P_e&=&\frac{\mu_e^4}{12\pi^2}+\frac{\mu_e^2T_e^2}{6}+\frac{7\pi^2T_e^4}{180}
-
\frac{e^2}{288}\left(5T_e^4+\frac{18\mu_e^2T_e^2}{\pi^2}+\frac{9\mu_e^4}{\pi^4}\right), 
\end{eqnarray}
whereas in the non-degenerate ultrarelativistic limit it reads:
\begin{eqnarray}\label{pdeg}
P_e&=&\frac{1}{16\pi^2}\left[p_F\mu_e(2\mu_e^2-m^2_e )-m^4_e\ln\left(\frac{\mu_e+p_F}{m_e}\right)\right]+\nonumber\\
&-&
\frac{e^2}{(2\pi)^4}\left\{\frac{3}{2}\left[\mu_e p_F-m_e^2\ln\left(\frac{\mu_e+p_F}{m_e}\right)\right]^2-p_F^4\right\}. 
\end{eqnarray}
In the latter case, 
the non-relativistic and relativistic limits will be considered in Sec.~\ref{sec4} for a degenerate plasma.

From the EoS for pressure, $P_e=P_e(\mu_e,T_e)$, the EoS for number density of electrons, $n_e=n_e(\mu_e, T_e)$, can be derived from the following thermodynamic relation: 
\begin{eqnarray}\label{n_e}
n_e=\frac{\partial P_e }{\partial \mu_e}.
\end{eqnarray}
Another relevant quantity for the description of the plasma is the energy density, $\epsilon_e=\epsilon_e(\mu_e,T_e)$, which reads:
\begin{eqnarray}\label{e_e}
\epsilon_e&=&-P_e+T_e\frac{\partial P_e}{\partial T_e}+\mu_e n_e.
\end{eqnarray}
These expressions for pressure, number density and energy density of the plasma will be important for our calculations when we incorporate crystal lattice effects.
This will be considered in the following.

\section{Crystalline features of stellar plasma}
\label{sec3} 

As we have previously mentioned, the outer crust of neutron stars and the core of white dwarfs have a crystal lattice structure~\cite{Haensel:2007yy,Chamel:2008ca,Drewsen}. 
This structure is generally called classical Coulomb crystal, which is set by the electromagnetic interactions between ions and electrons in the stellar plasma. At low temperatures,
this crystal is well known in solid state physics  and can be reproduced experimentally in Earth laboratory~\cite{quantumcrystal,Drewsen}. 
The crystallization process is governed by the parameter $\Gamma_m$, also known as Coulomb parameter~\cite{quantumcrystal}:
\begin{eqnarray}\label{cristalizacao}
\Gamma_m=\frac{e^2Z^2}{k_{B}R_{\rm cell} T_m},
\end{eqnarray} 
where
\begin{eqnarray}\label{raiocell}
R_{\rm cell}=\left(\frac{4 \pi n_e}{3 Z}\right)^{-1/3}. 
\end{eqnarray}
This process depends directly on the charge number of lattice ions, $Z$, the radius of the WS cell, $R_{\rm cell}$, 
and the melting temperature, $T_m$,
where
$R_{\rm cell}$ depends on the electron number density, Eq. \eqref{raiocell}.
The crystal phase occurs for $T<T_m$, whereas for $T>T_m$ the ions can move freely, and the substance becomes a liquid.
The dependence on these parameters allows flexibility in the configuration of the lattice ions.
From simulations~\cite{Drewsen,simulations}, one obtains the lower limit for $\Gamma_m$,  $\Gamma_m \gtrsim 173$, for the OCP.

However, for a stellar scenery the  plasma characteristic depends on its depth in the stellar structure. As we go deeper from the surface towards the center along its radius, energy and temperature increase and, we have greater ionization, until we reach full ionization (bare ions). 
Deeper yet in the plasma, there are a few free neutrons (order of $1\% $), which are beyond the scope of this work. 
 
Coulomb crystals are possible in both OCP and MCP cases. 
In the former, one single species are the lattice ions, which coexists with the electron background,
whereas in the latter multiple ion species are present. 
Additionally, in deeper regions of the outer crust of neutron stars mix ion crystals can occur,
which would demand a MCP model in order to describe the system. 
The OCP and MCP models that include quantum effects are called quantum Coulomb crystal or Yukawa crystal~\cite{quantumcrystal}.

The ground state of the OCP of lattice ions in the stellar medium corresponds to a bcc lattice.
Particularly, in a OCP crystal at high densities, we expect the formation of the bcc lattice as well as other types of lattices, such as the face-centered-cubic (fcc) and hexagonal-close-packed (hcp). 
The simple Coulomb crystal is unstable for a cubic configuration (See Sec.~2.3.3 of Ref.~\cite{Haensel:2007yy} for more details).

In the following we introduce the bcc lattice structure in the QED plasma using linear response theory. 
We will consider a bcc lattice with only one type of ion, although we present a model that allows an analysis with a mixed bcc crystal lattice (mixed lattices can be found in deeper regions of the outer crust of neutron stars~\cite{Chamel:2008ca}).

\section{Crystal Lattice Effects in the QED Plasma}
\label{sec4}

We consider a crystal lattice of ions immersed in the QED plasma, where the Coulomb interaction affects the whole charge distribution of the electron plasma and creates a non homogeneous distribution.
A static lattice is assumed for simplicity (Bravais lattice), i.e., we neglect lattice vibrations, assuming that its presence does not introduce extra interactions except for an external classical potential.
Also, despite the temperature of the plasma and of the ions are different, the plasma temperature is dominant and we neglect the effect of the lattice temperature (see Ref.~\cite{Haensel:2007yy}). 
We adopt the bcc lattice due to its stability.
We consider that the fundamental state of the ions in the stellar medium corresponds to the bcc lattice, both for OCP and for MCP.

The interaction between the lattice ions and the electron plasma is done by considering that: (i) the ions are classic due to the mass difference compared to the electrons; (ii) the interaction is only Coulombian. For the lattice set up, we consider a cubic lattice with charge $Z_1 e$ and another cubic lattice with charge $Z_2 e$, where $e$ is the electron charge, whose centers are displaced by $a/2$ in each spatial direction, where $a$ is the primitive lattice parameter of the  Bravais lattice.
Due to the geometric construction of the WS cell of a bcc lattice,
the relation between the primitive lattice parameter  and the radius of WS cell is given by $R_{\rm cell}=a/2$, and from now on the primitive lattice parameter will be written only in terms of $R_{\rm cell}$. 

The charge density of the lattice can be written in the form:
\begin{eqnarray}\label{redeclassica}
\rho_{\rm cl}(\vec{r})= 
e
\sum_{n_{r}}
\left[
Z_1\delta_{n_{r}\,a}(\vec{r})
+
Z_2\delta_{\left(n_{r}\,+1/2\,\hat{r}\right)a}(\vec{r}) \right],
  \end{eqnarray}
where $\delta_{r_0}(\vec{r})=\delta(\vec{r}-\vec{r}_0)$,
$\hat{r}$ is the radial unit vector and $n_{r}=(n_x,n_y,n_z)$ is the vector of neighbors counting factors, $n_x,n_y,n_z\in{\mathbb Z}$. 
For a bcc lattice, the charge numbers $Z_1$ and $Z_2$  can assume different values. 
This charge configuration is adopted because it is possible to have the presence of binary mixtures of nuclides in the region of the crust, composing a mixed lattice.
For Eq.~\eqref{redeclassica} we do not consider electron penetration in the nuclei and neglect the effects of magnetic fields. 
We only take into account the nucleus radius, which is written as $R_0 = 1.2 \times A^{1/3}\, {\rm fm} $.
 
The understanding of Coulomb crystal permits to establish the  relationship between its structure parameters with temperature and number density. However, in this work an alternative method to track lattice formation is derived using quantum field theory, which encompasses both quantum and relativistic regimes.
The calculation in the method is detailed  in Appendix~\ref{appendixrcellsec}.
It involves an arbitrary number of neighbor ions, differently of the methods considered in Refs.~\cite{Chamel:2008ca, Rotondo:2011zz, Ruester:2005fm}, where only the influence of the first nearest neighbor ions are taken into account.

We have considered that each electron of the plasma interacts electrically with all the crystal lattice ions. 
From the point of view of a single cell, we can identify a first nearest neighborhood of ions, as well a second, a third and so on. The neighborhood level is taken with respect to the WS cell center.
The nearest ones are responsible for the leading order contributions for the electric potential energy, whereas the more distant ones are next to leading order corrections.

To analyse the effects of the crystal lattice in the QED plasma, we adopt the linear response theory in the context of quantum field theory~\cite {kapusta, bellac, walecka}. 
The plasma is described as a free electron gas subject to electron exchange interaction corrections. 
We consider the electric field produced by the ions as an external source by means of linear response theory. 
For this purpose, we must consider a linear interaction between the electromagnetic plasma field, described by the four-vector potential $A_{\mu}$, and a classical current $j_{\rm cl}^\mu$, which describes the lattice ions as an external source. 
This interaction term reads:
\begin{eqnarray}\label{Hext}
\hat{H}_{\rm ext}(t)
=
-\int d^3x\,
j_{\rm cl}^\mu(\vec{x},t)\, \hat{A}_{\mu}(\vec{x},t),
\end{eqnarray}
where $\hat{H}_{\rm ext}$ is the Hamiltonian of the external source and the classical source satisfies the covariant Maxwell equations $\partial_\nu F^{\mu\nu}_{\rm cl}=j_{\rm cl}^\mu$.
The linear response in $\hat{A}_{\mu}$ due to the effect of the external field can be written as:
\begin{eqnarray}\label{deltaA}
\delta\langle \hat{A}_{\mu}(\vec{x},t)\rangle
&=&
i\int dt'\, {\rm Tr}\left\{\hat{\rho}[\hat{H}_{\rm ext}(t'),\hat{A}_{\mu}(\vec{x},t)]\right\}\theta(t-t')\nonumber\\
&=&
\int dt'\int d^3x'\, 
D^{\rm R}_{\mu\nu}(\vec{x}-\vec{x}',t-t')
\,
j_{\rm cl}^\nu(\vec{x}',t')
,
\end{eqnarray}
where $\delta\langle\hat{A}_{\mu}\rangle$ is the average of the field $A_{\mu}$
and
$D^{\rm R}_{\mu\nu}(\vec{x}-\vec{x}',t-t')=i\,{\rm Tr}\left\{\hat{\rho}[\hat{A}_{\mu}(\vec{x},t),\hat{A}_{\mu}(\vec{x}',t')]\right\}\theta(t-t')$ is the retarded photon propagator.
From  $\langle\hat{A}_{\mu}\rangle_{\rm eq}=0$, one notices that $\delta A_\mu=A_\mu$.
In the momentum space, from Eq.~\eqref{deltaA}, the field $A_{\mu}$ reads:
\begin{eqnarray}\label{Amu}
A_{\mu}(\vec{x},t)
=
i
\int 
\frac{d\omega}{2\pi}
\frac{d^3q}{(2\pi)^3}
e^{i(\vec{q}\cdot\vec{x}-\omega t)}
D_{\mu\nu}^{\rm R}(q,\omega)
j^{\nu}_{\rm cl}(q,\omega).
\end{eqnarray}
The retarded photon propagator can be decomposed as:
\begin{eqnarray}\label{Amu2k}
D_R^{\mu\nu}(K)
=
\frac{P^{\mu\nu}_T}{G-K^2}+\frac{P^{\mu\nu}_L}{F-K^2}+\frac{\xi\,K^{\mu}K^{\nu}}{K^4},
\end{eqnarray}
where $G=G(\vec{k},\omega)$ and $F=F(\vec{k},\omega)$ are scalar functions and $P^{\mu\nu}_T$ and $P^{\mu\nu}_L$ are the transverse and longitudinal projectors, respectively. 
These functions are properly defined in Refs.~\cite{bellac,kapusta}.

The gauge field can be explicitly written as $A^{\mu}=(\Phi,\vec{A})$, where $\Phi$ is the scalar electric potential and $\vec{A}$ is the vector potential. 
We are interested in the calculation of $A^{0}$, since we are not considering any magnetic external source.
Additionally, we want to obtain the lattice potential energy accounting for the plasma effects, which stems from $A^{0}(\vec{x})$.

Choosing the covariant Landau gauge, $\xi=0$, and considering the explicit expressions and properties of $P^{\mu\nu}_T$ and $P^{\mu\nu}_L$, from Eq.~\eqref{Amu2k} for the $\mu=0$ component, one obtains:
\begin{eqnarray}
A^0(K)=-\frac{\rho_{\rm cl}(K)}{(k^0)^2-\vec{k}^2-F(\vec{k},\omega)}.
\end{eqnarray}
The inverse of Fourier transform of the latter equation reads:
\begin{eqnarray}
A^0(x)=-\int \frac{d^4K}{(2\pi)^4}e^{-iK_{\mu}x^{\mu}}\frac{\rho_{\rm cl}(K)}{(k^0)^2-\vec{k}^2-F(\vec{k},\omega)}. \label{Ax0}
\end{eqnarray}
The charge density in the momentum space is given by the Fourier transform:
\begin{eqnarray}\label{rhoclK}
\rho_{\rm cl}(K)=\int d^4x e^{iK_{\mu}x^{\mu}}\rho_{\rm cl}(\vec{x}).
\end{eqnarray}
From Eq.~\eqref{redeclassica}, the latter equation reads:
\begin{eqnarray}\label{rhocl}
\rho_{\rm cl}(K)=2\pi e \sum_{n_{r}}\left[Z_1 e^{ia\vec{k}\cdot n_{r}}+ Z_2e^{ia\vec{k}\cdot\left(n_{r}+\frac{1}{2}\hat{r}\right)}\right]\delta(k^0).
\end{eqnarray}
Substituting Eq.~\eqref{rhocl} in Eq.~\eqref{Ax0}, the result reads:
\begin{eqnarray}
A^0(x)=e\sum_{n_{r}}\int\frac{d^3\vec{k}}{(2\pi)^3}e^{i\vec{k}\cdot(\vec{r}+a\,n_{r})}\left(
\frac{Z_1  +Z_2 e^{i\frac{a}{2}\vec{k}\cdot \hat{r}}}{\vec{k}^2+F(\vec{k},0)}\right), \label{A0P}
\end{eqnarray}
where we have set $\omega=0$ in $F(\vec{k},\omega)$ due to the static nature of the external perturbation, i.e., the static lattice of ions.

Equation~\eqref{redeclassica} is an oversimplified description for our scenario, since it describes point-like ions in the lattice. However, the nucleus has a finite radius and, additionally, electrons penetrate the nuclei volume. Thus, it is necessary to consider both effects for the interaction of the electron gas with the nuclei charges. For this purpose we correct the charge number $Z$ in Eq.~\eqref{redeclassica} as an effective charge, $Z_{\rm eff}$, which is obtained by~\cite{stella},
 \begin{eqnarray}\label{zeff}
Z_{\rm eff}=Z-\int_0^{R_0}4\pi r^2n_e(\bar{\mu}_{\rm  in}(r))dr,
\end{eqnarray}
where $n_e(\bar{\mu}_{\rm in}(r))$ and $\bar{\mu}_{\rm in}(r)$ are the electron number density and the effective chemical potential, respectively, inside the nucleus.
It is important to point out that the effective chemical potential inside the nucleus contains three different contributions:
(i) the first one comes from the homogeneous chemical potential $\mu_e$;
(ii) the second one comes from the electron interactions with the screened charged ions outside the nuclear volume, described in terms of Eq.~\eqref{A0P};
(iii) the third one comes from the electron interactions with the nuclear charge distribution~\footnote{We have considered the electric potential from Ref.~\cite{flambaum} for the nucleus.}, approximately considered as uniform inside the spherical nucleus. 

From these considerations and from Eq.~\eqref{zeff}, we rewrite Eq.~\eqref{A0P} as:
\begin{eqnarray}\label{suma0r}
A^0(\vec{r})
=
\sum_{n_{r}}\sum_{i=1}^{2}
\frac{e Z_i^{\rm eff} }{2\pi^2 |\vec{r}-\vec{r}_i|}\int_{0}^{\infty}\frac{dk  \,k\sin{(k\,|\vec{r}-\vec{r_i}|)}}{k^2+F(0,\vec{k})}.
\end{eqnarray}
We set $Z_i=Z_i^{\rm eff}$, $(i=1,2)$, and $r_i$ are the ions positions, given by:
\begin{eqnarray}\label{ri}
\vec{r}_1&=&n_x\,a\hat{i}+n_y\,a\hat{j}+n_z\,a\hat{k},\\
\vec{r}_2&=&\left(n_x\,a+\frac{a}{2}\right)\hat{i}+\left(n_y\,a+\frac{a}{2}\right)\hat{j}+\left(n_z\, a+\frac{a}{2}\right)\hat{k}.
\end{eqnarray}

Thus, the result given by Eq.~\eqref{suma0r} is the electrostatic potential of the lattice ions screened by the QED plasma.
Multiplying this potential by the electron charge $e$, we obtain the variation of the free energy of the system~\cite{kapusta}, which is the electrical potential energy between the lattice ions and the electrons, given as:
\begin{eqnarray}\label{v}
U(\vec{r})=eA^0(\vec{r}).
\end{eqnarray}
This equation adapts Eq.~(3.4) given in Ref.~\cite{Kapusta:1988fi}, originally for interacting charges in a QED plasma medium, which in this study is the interaction between lattice ions and electrons with sized ions.
The presence of lattice ions in the plasma induces a variation in the chemical potential of the homogeneous system, which is now non homogeneous.
This chemical potential perturbation, $\delta\mu(\vec{r})$, due to the electrical potential of the ions, will be an additional contribution to the chemical potential of the plasma~\cite{han}.
This contribution reads:
\begin{eqnarray}\label{deltamu}
\delta\mu(\vec{r})=eA^0(\vec{r}).
\end{eqnarray}
 Therefore, the plasma chemical potential is now an electrochemical potential and position dependent, defined as:
\begin{eqnarray}\label{mubar}
\bar{\mu}_e(\vec{r})=\mu_e+\delta\mu(\vec{r}).
\end{eqnarray}
However, we will simply call it effective chemical potential.
This is how the lattice ions affect the plasma EoS. 
Local hydrostatic equilibrium is satisfied as one can see in Appendix~\ref{equilibriumappendix}.

Our main goal is to calculate the EoS of the plasma in the presence of a crystal lattice.
This is done conciliating the EoS and the electric potential, Eq.~\eqref{v}.
In the following, we present the EoS including lattice effects for the degenerate gas limit. 
The non-degenerate ultra-relativistic limit was not considered here since the crystal lattice formation was not found within our approach.
However, we cannot rule out the formation of a crystal lattice for a different configuration.

\section{Thermodynamic Quantities inside the WS Cell}
\label{sec5}

In this section, 
we introduce the corrections for pressure, number density and energy density of the plasma when a bcc lattice is immersed on it. Local values for these quantities are determined inside the WS cell of the periodic lattice. 

To illustrate the calculation  scheme adopted for the correction in thermodynamic quantities inside the WS cell, the present work take two particular situations of the electron plasma of great importance in dealing with white dwarf and neutron star crust structure. Namely, the non-relativistic degenerate gas for the first and relativistic degenerate gas for the second. 
The main correction consists in substituting 
the chemical potential of electrons by the correspondent effective chemical potential defined in Eq.~\eqref{mubar}.

The non-relativistic degenerate gas regime ($T_e=0$ and $p_F\ll m_e$) corresponds to the outermost region of cold white dwarfs. 
For example, the EoS of pressure, Eq.~\eqref{pdeg}, reduces to:
\begin{eqnarray}\label{pdegmu}
P_e(\bar{\mu}_e)
=\frac{p_F(\bar{\mu}_e)^5}{30\pi^2m_e}
\left(1+\frac{15e^2}{8\pi^2}\frac{m_e}{p_F(\bar{\mu}_e)}\right),
\end{eqnarray}
where $p_F(\bar{\mu}_e)=\sqrt{2m_e(\bar{\mu_e}-m_e)}$.
Numerical calculations of $A^0$, Eq.~\eqref{suma0r}, are performed in order to obtain explicitly the effects of the crystal lattice in the EoS of pressure.
From the result of Eq.~\eqref{pdegmu}, we can obtain the number and energy densities, given by Eqs.~\eqref{n_e} and~\eqref{e_e}, respectively.

The relativistic degenerate gas scenery ($T_e=0$ and $p_F\gg m_e$) corresponds to the high density region of the neutron star crust and white dwarf core until it has completely cooled down.
Analogously, the EoS of pressure, Eq.~\eqref{pdeg},  results:
\begin{eqnarray}
P_e(\bar{\mu}_e) &=&
\frac{\bar{\mu}_e^4}{12\pi^2}\left(1-\frac{3e^2}{8\pi^2}\right),
\end{eqnarray}
where $\bar{\mu}_e (\vec{r})$ is obtained by the numerical calculation of $A^0(\vec{r})$ in this limit.

It is necessary to be cautious when choosing ion charge values, since fully ionized ions can be found at mass densities 
$\rho\,\gg\,11\,AZ\,{\rm g/ cm}^{3}$ (see Refs.~\cite{Haensel:2007yy, Ruester:2005fm}). 
In the relativistic case, the atoms are fully ionized, whereas in the non-relativistic case the atoms can be both partially or fully ionized. 
However, in this work we only consider atoms which are completely ionized.

\section{Thermodynamic Quantities in the hydrodynamic scale}
\label{sec6}

In this section, we discuss how to define thermodynamics quantities at hydrodynamic level in order to establish the EoS of the plasma in stellar medium.
 These EoS can be used for applications to determine the structure of compact objects. 
At the hydrodynamic scale, which we are considering as the macroscopic level, it is necessary to consider the response of each cell as a whole, taking the cell as the infinitesimal element of the stellar structure.
For this purpose, we need to define effective (macro) values of thermodynamic variables for each cell associated to the medium as a whole.

\begin{figure}[htb!]
\centering 
\includegraphics[width=5cm]{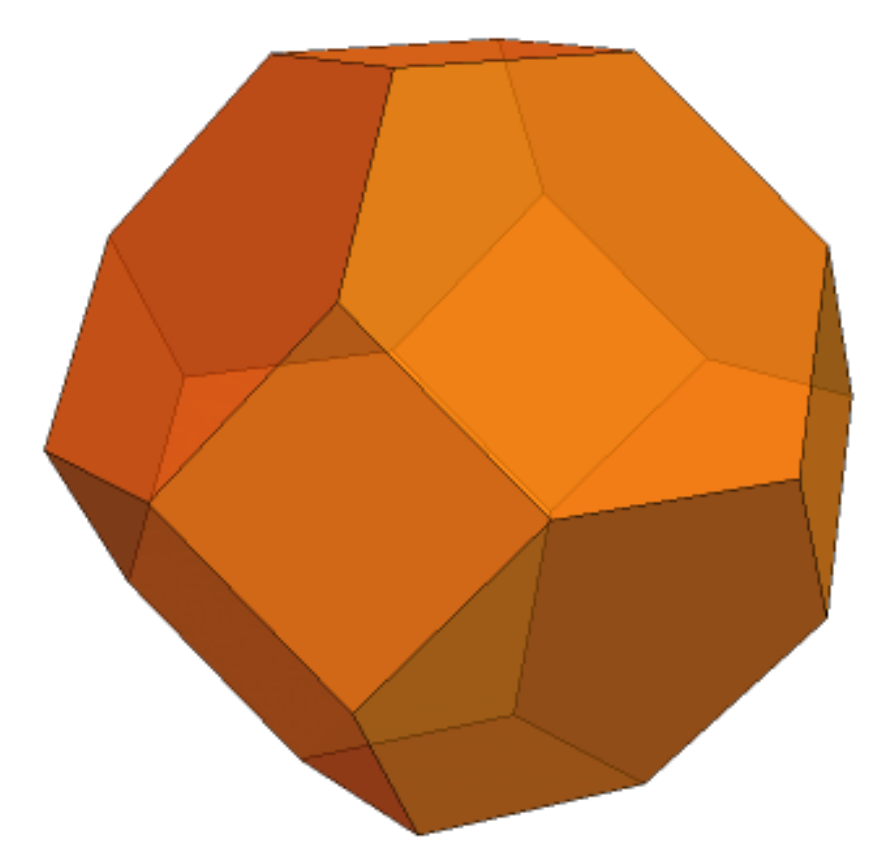} \caption{Wigner-Seitz cell in the body-centered cubic crystal lattice, whose shape is a truncated octahedron.}
\label{truncated_octahedron}
\end{figure}

Indeed, the actual form of the WS cell is a polyhedral one, which is illustrated in Fig.~\ref{truncated_octahedron}. However, in the literature the WS cell is usually approximated by a spherical form (see Ref.~\cite{Chamel:2008ca}). 
The radius of WS cell is determined in the Appendix~\ref{appendixrcellsec} by calculating the contribution of arbitrary neighbor ion interactions.

The macro pressure $P$ associated to the WS cell can be defined by the pressure value on the cell surface in the spherical approximation following Ref.~\cite{Rotondo:2011zz}.
On the other hand, the macro mass density is defined as the total mass of the cell divided by the cell volume, where the mass of the cell is obtained here from the Bethe-Weizs\"acker semi-empirical mass formula~\cite{chung}. 
The macro energy density is defined as total energy of a single cell, and is calculated using the relation between energy and pressure given by the relativistic relation $E = \rho c^2 +(1/V_{\rm WS})\int_{V_{\rm WS}}\epsilon_e dV$.
This is an important quantity to solve the Tolman-Oppenheimer-Volkoff (TOV) structure  equations~\cite{Tolman:1939jz,Oppenheimer:1939ne}.

\section{Results and discussion}
\label{result}

In this section we present numerical results for the EoS of pressure, number density and energy density for a non homogeneous electron gas for one of the scenarios discussed in Sec.~\ref{sec5}, where the non-homogeneity is due to the presence of the crystal lattice.

The analysis is performed for both the interior of a single primitive cell and for the macro values of thermodynamic variables associated to each WS cell.  
For the interior of primitive cells, we provide the analysis of the influence of the crystal lattice in the electron distribution.
On the other hand, the extended definitions of macro thermodynamics variables show the relevance of the ion-electron interaction in the crystal and should be used for applications in the structure of compact objects.
Consequently, both the EoS of the electron gas and the WS radius are calculated in a  consistent way.

We have used beta decay and electron capture data in order to obtain threshold mass density values of these process and compare our results with other models of the literature.
We have considered that the lattice ions do not change due to nuclear decay or electron capture.

\subsection{Thermodynamic variables distribution inside Wigner-Seitz cell}
\label{results_interior}

The purpose of this section is to illustrate how physical quantities such as pressure, number and energy densities, and chemical potential are modified due to the presence of ion lattice in the plasma. 
We present some plots that illustrate one of the scenarios presented in Sec.~\ref{sec5} as a representative case, which  is the relativistic degenerate gas. 
From this analysis, we have established the influence of the presence a crystal lattice in the plasma.

\begin{figure*}[htb!]
\centering
\subfigure[]{
\includegraphics[width=0.45\textwidth]{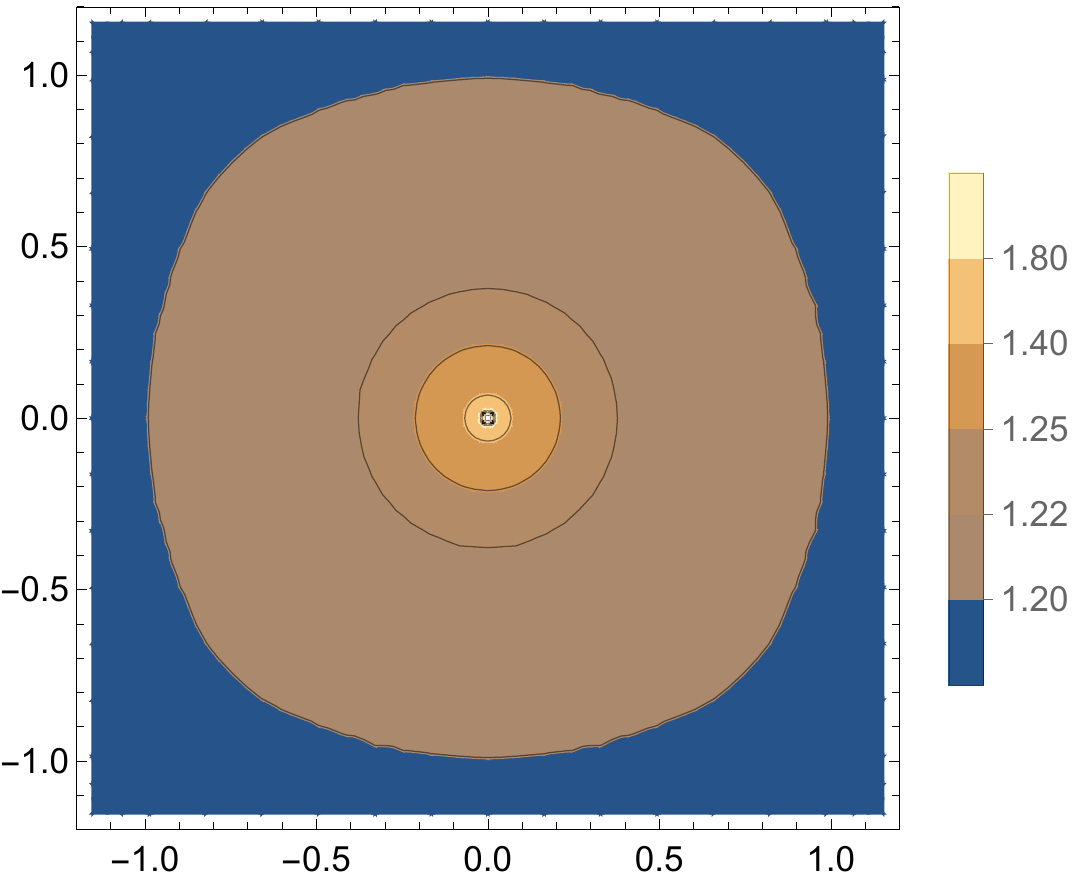}
}         \qquad
\subfigure[]{
\includegraphics[width=0.45\textwidth]{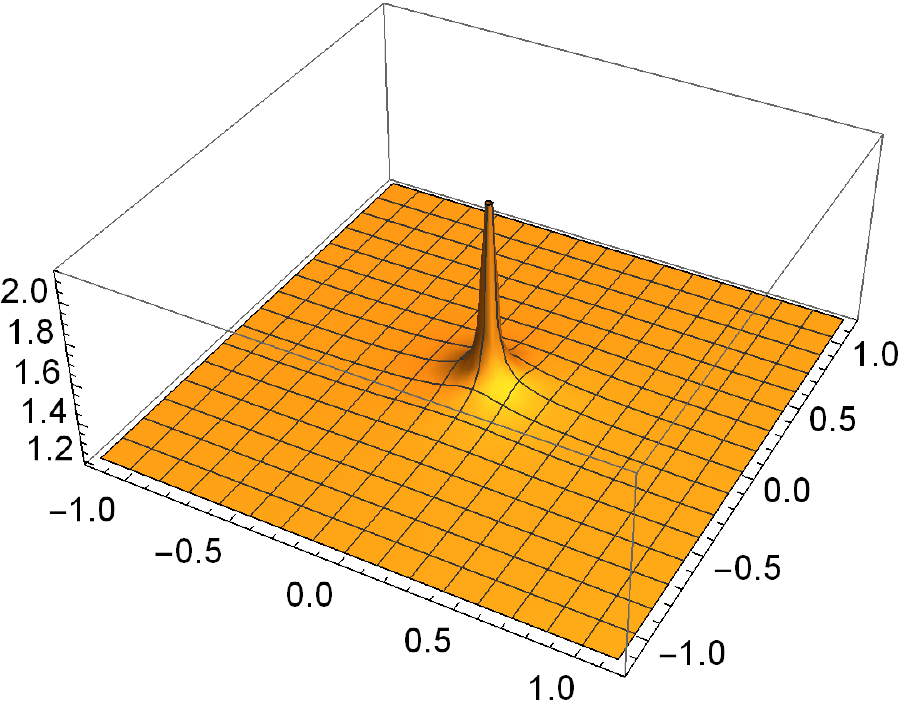}
}       \qquad
\\
\subfigure[]{
\includegraphics[width=0.45\textwidth]{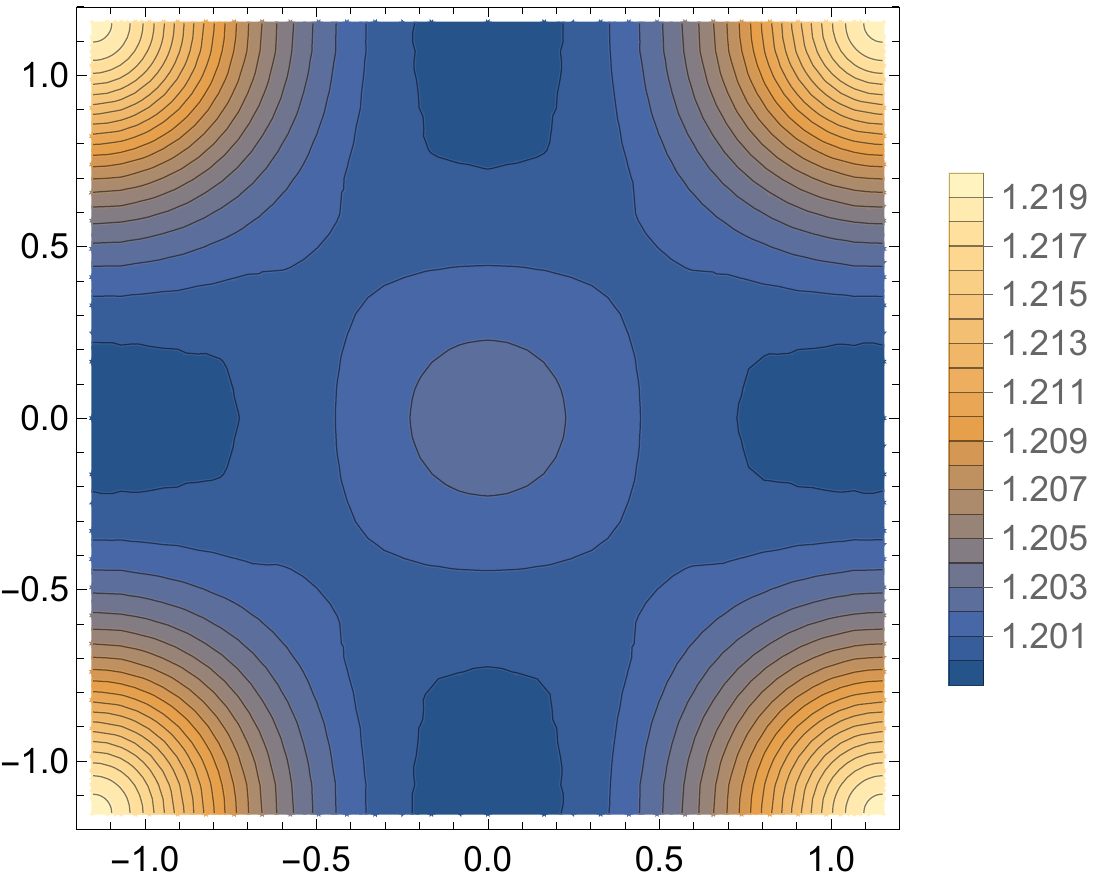}
}      \qquad
\subfigure[]{
\includegraphics[width=0.45\textwidth]{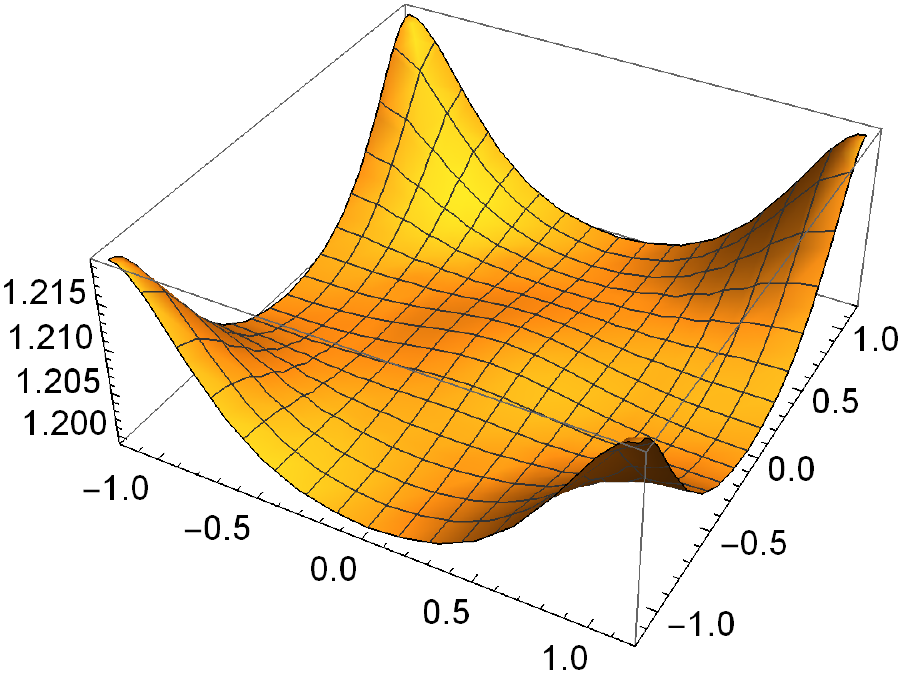}}
\caption{
Ratio between non homogeneous and homogeneous chemical potentials, $\bar{\mu}_e$ and ${\mu}_e$, respectively,  for different cross-sections.
We consider the following representative values: number of neighbors $n = 10$, charge numbers $Z_1=Z_2=8$, homogeneous chemical potential $ \mu_e = 4\, {\rm MeV}$, atomic mass $A=16$, and WS radius $ R_{\rm cell} = 1.151092\, {\rm MeV}^{-1}$.
Panels [a] and [b] represent 
$\bar{\mu}_e(x, y, 0)/\mu_e$,
whereas
panels [c] and [d] represent
$\bar{\mu}_e(x, y, a/3)/\mu_e$.} 
\label{fig:deghigh_mu}
\end{figure*}

\begin{figure*}[htb!]
\centering\subfigure[]{
\includegraphics[width=0.45\textwidth]{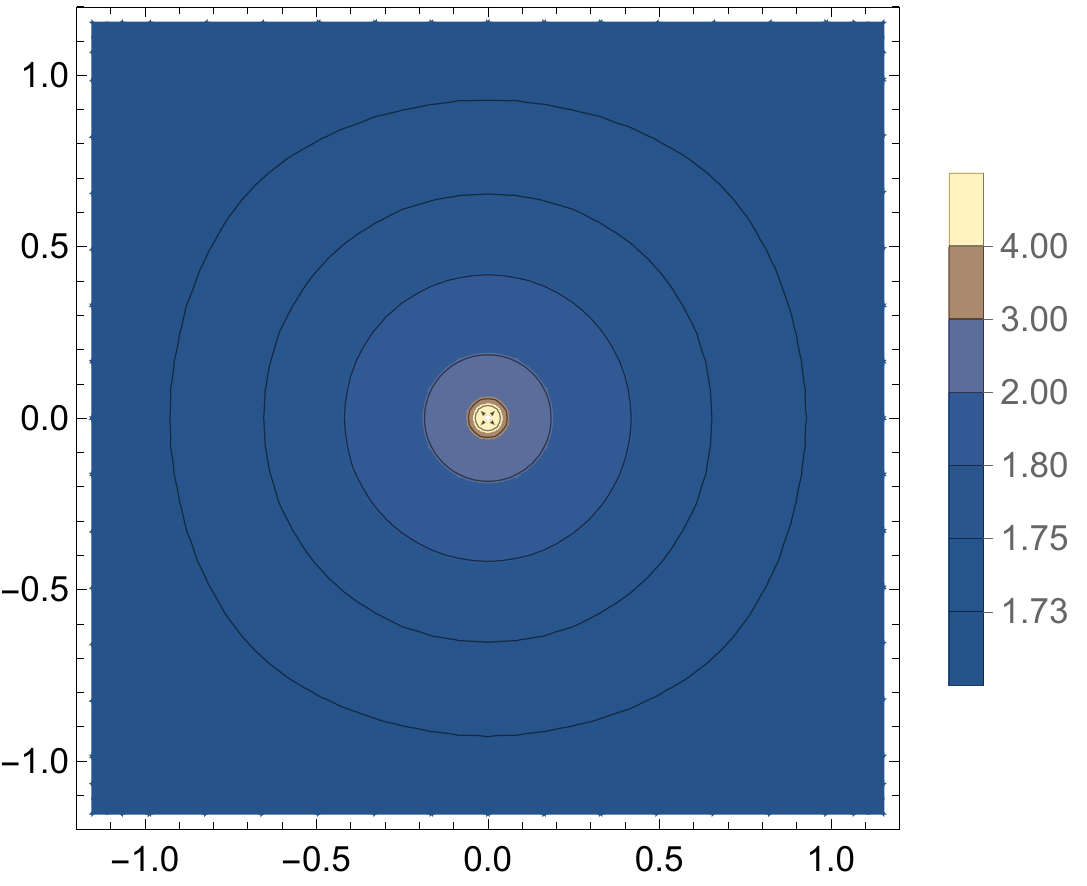}
}         \qquad
\subfigure[]{
\includegraphics[width=0.45\textwidth]{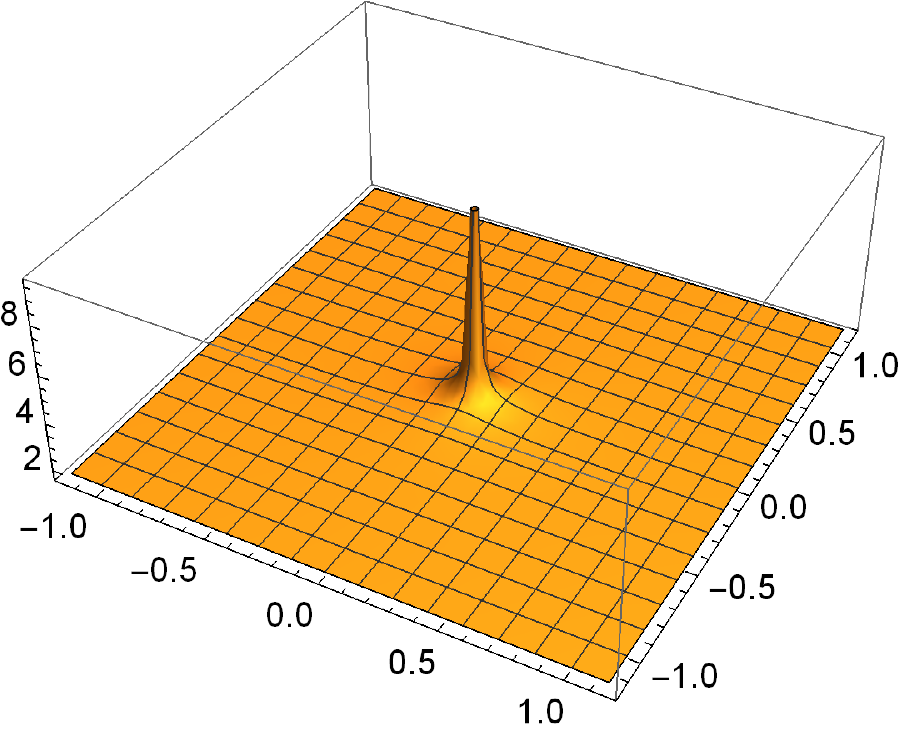}
}       \qquad
\\
\subfigure[]{
\includegraphics[width=0.45\textwidth]{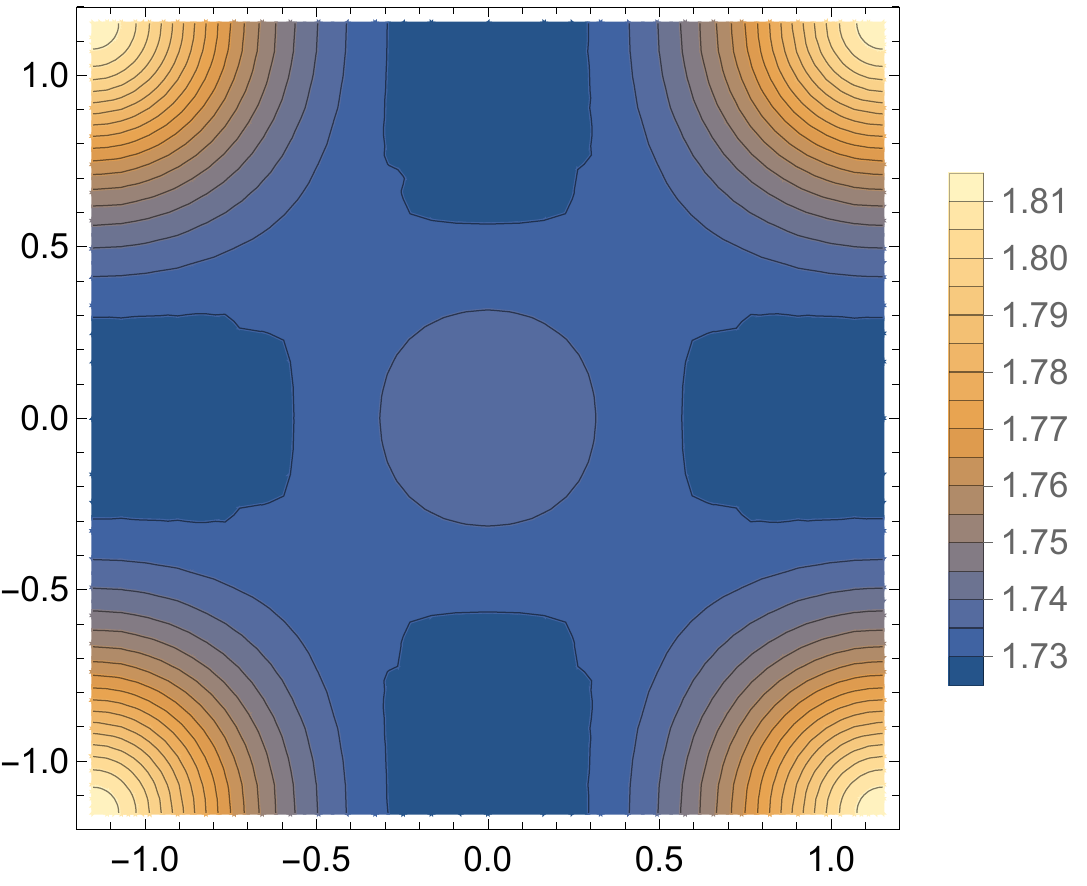}
}      \qquad
\subfigure[]{
\includegraphics[width=0.45\textwidth]{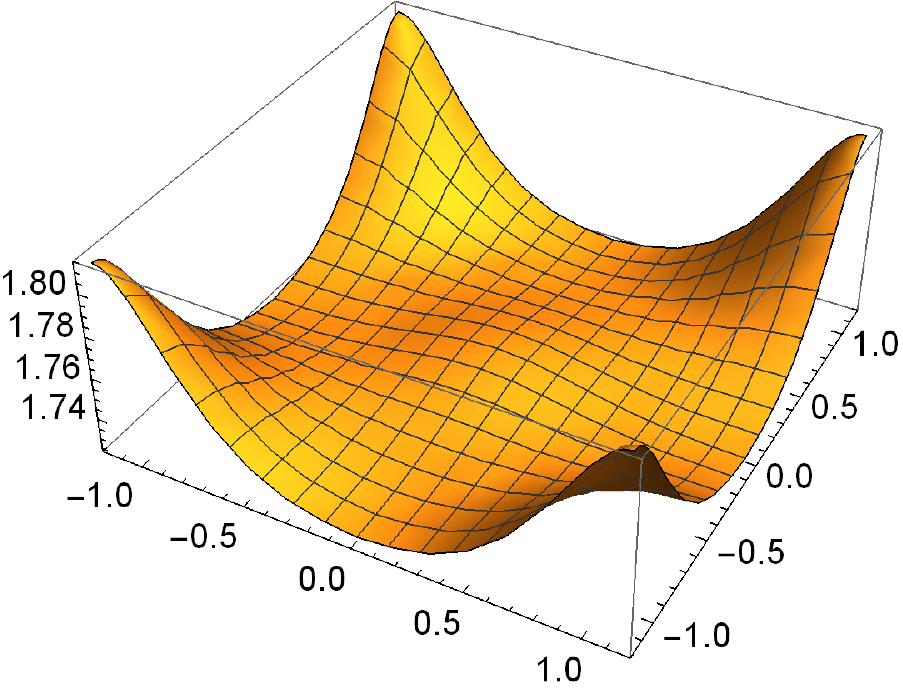}}
\caption{
Ratio between non homogeneous and homogeneous number density of electrons, $\bar{n}_e$ and ${n}_e$, respectively,  for different cross-sections.
We consider the following representative values: number of neighbors $n = 10$, charge numbers $Z_1=Z_2=8$, homogeneous chemical potential $ \mu_e = 4\, {\rm MeV}$, atomic mass $A=16$, and WS radius $ R_{\rm cell} = 1.151092\, {\rm MeV}^{-1}$.
Panels [a] and [b] represent
$\bar{n}_e(x, y, 0)/{n}_e$,
whereas
panels [c] and [d] represent
$\bar{n}_e(x, y, a/3)/{n}_e$.}  \label{fig:deghigh_n}
\end{figure*}

\begin{figure*}[htb!]
\centering\subfigure[]{   \includegraphics[width=0.45\textwidth]{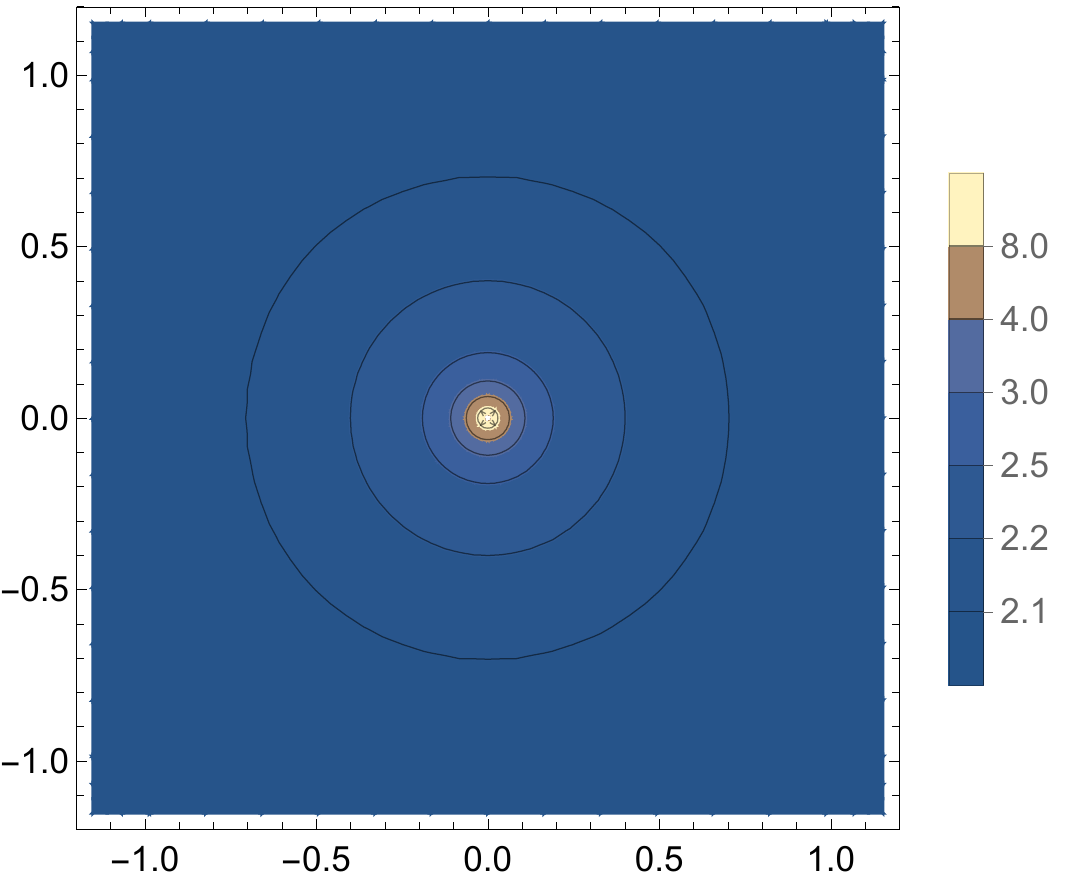}
}         \qquad
\subfigure[]{
\includegraphics[width=0.45\textwidth]{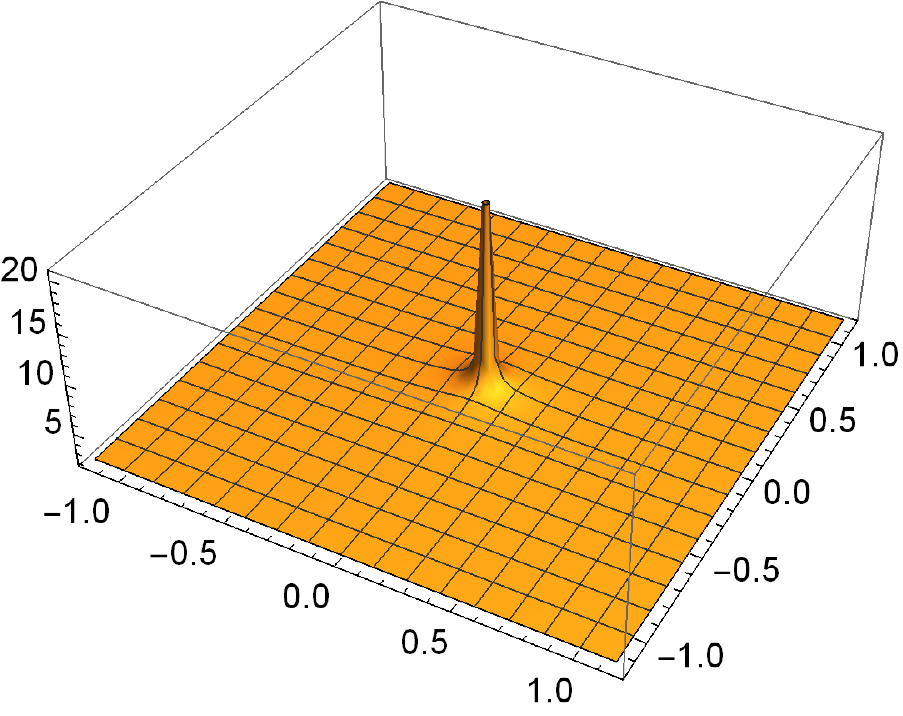}
}       \qquad
\\
\subfigure[]{
\includegraphics[width=0.45\textwidth]{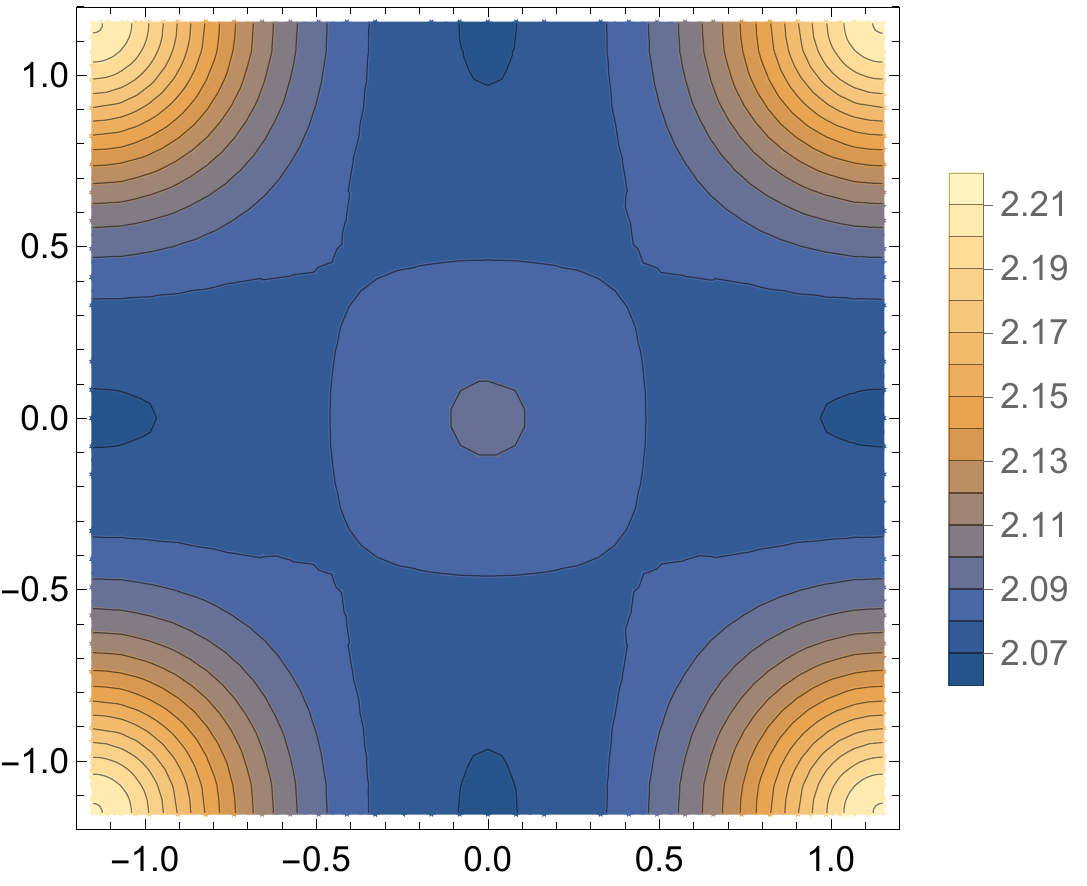}
}      \qquad
\subfigure[]{
\includegraphics[width=0.45\textwidth]{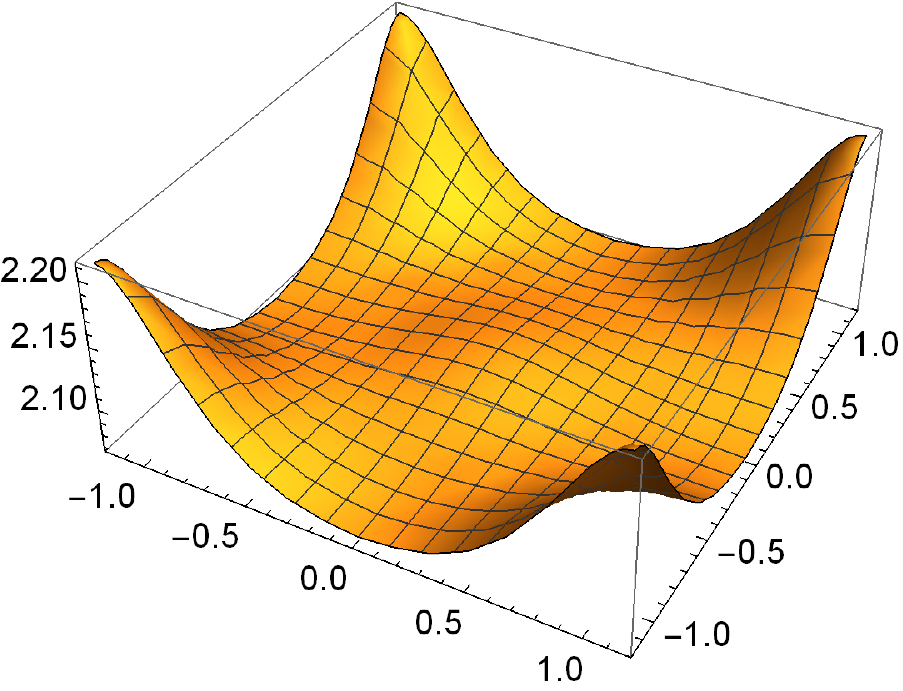}}
\caption{
Ratio between non homogeneous and homogeneous pressure of electrons, $\bar{P}_e$ and ${P}_e$, respectively,  for different cross-sections.
We consider the following representative values: number of neighbors $n = 10$, charge numbers $Z_1=Z_2=8$, homogeneous chemical potential $ \mu_e = 4\, {\rm MeV}$, atomic mass $A=16$, and WS radius $ R_{\rm cell} = 1.151092\, {\rm MeV}^{-1}$.
Panels [a] and [b] represent 
$\bar{P}_e(x, y, 0)/{P}_e$,
whereas
panels [c] and [d] represent
$\bar{P}_e(x, y, a/3)/{P}_e$.}  \label{fig:deghigh_P}
\end{figure*}

In order to have a more deep view of the crystal lattice effect in the plasma, we plot the ratio between non homogeneous and homogeneous variables, where the non homogeneous are identified with an overbar. 
In Figures.~\ref{fig:deghigh_mu},~\ref{fig:deghigh_n} and~\ref{fig:deghigh_P}, 
we consider these ratios for the chemical potential, number density of and pressure of electrons, respectively.
We chose the values $Z_1=Z_2=8$ for the charge number, $R_{\rm cell}=1.151092\, {\rm MeV}^{-1}$ for the WS cell radius, $\mu_e=4\, {\rm MeV}$ for the chemical potential and $A=16$ for the atomic number.
Each figure has four panels, which consider different cross-sections of the WS cell.
In each figure, panels (a) and (c) show contour lines of intensity of the ratios for each thermodynamic variable,
whereas panels (b) and (d) are 3D plots of these ratios.

\subsection{Macro values for thermodynamic variables associated to a single Wigner-Seitz cell} 
\label{results_exterior}

Once again we consider the particular case of  the degenerate relativistic situation.
However, we now point out this case as our main results due to its   applications for compact objects. 
The scenario describes the physics of most part of the outer crust of neutron stars and of the core of white dwarfs. 

The method presented in Appendix~\ref{appendixrcellsec} shows that the neighborhood order affects the value of $R_{\rm cell}$, which means that to consider only the first neighborhood may compromise the precision of calculations.
We have performed our numerical analysis considering contributions far beyond the first neighborhood.

The main objective is to obtain the macro pressure as a function of mass density and the energy density as a function of pressure, namely, the EoS of the plasma. 
As presented in Sec.~\ref{sec6}, the macro pressure $P$ of the WS cell is defined as the micro pressure of the cell interior on the boundary of the cell, whereas the energy is defined as total energy of a single cell.

\begin{figure}[htb!]
\centering
\includegraphics[width=10cm]{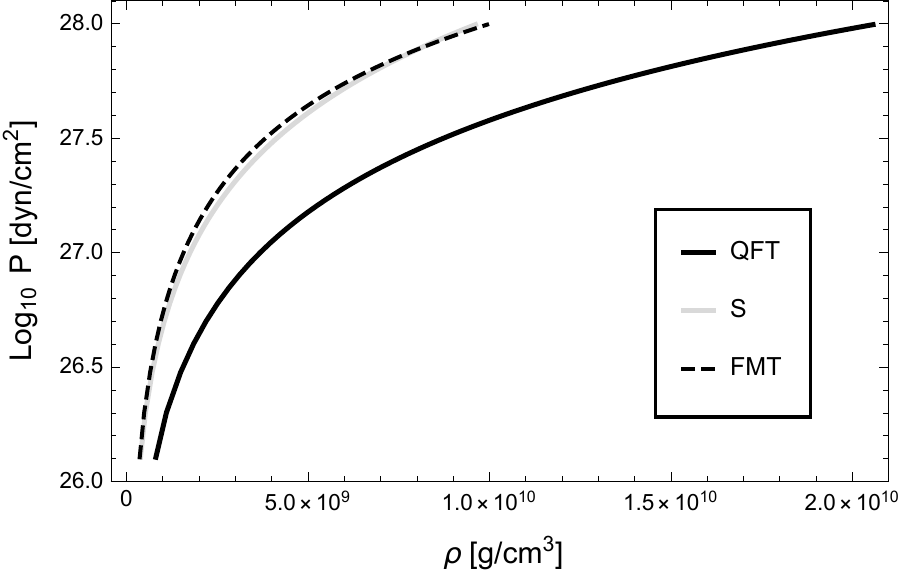} \caption{Comparison of EoS $P(\rho)$ for ${\rm C}^{12}_{6}$ between the Salpeter (S), Feynman-Metropolis-Teller (FMT) and the present model using quantum field theory (QFT).}
\label{densvspressao}
\end{figure}

\begin{figure}[htb!]
\centering 
\includegraphics[width=10cm]{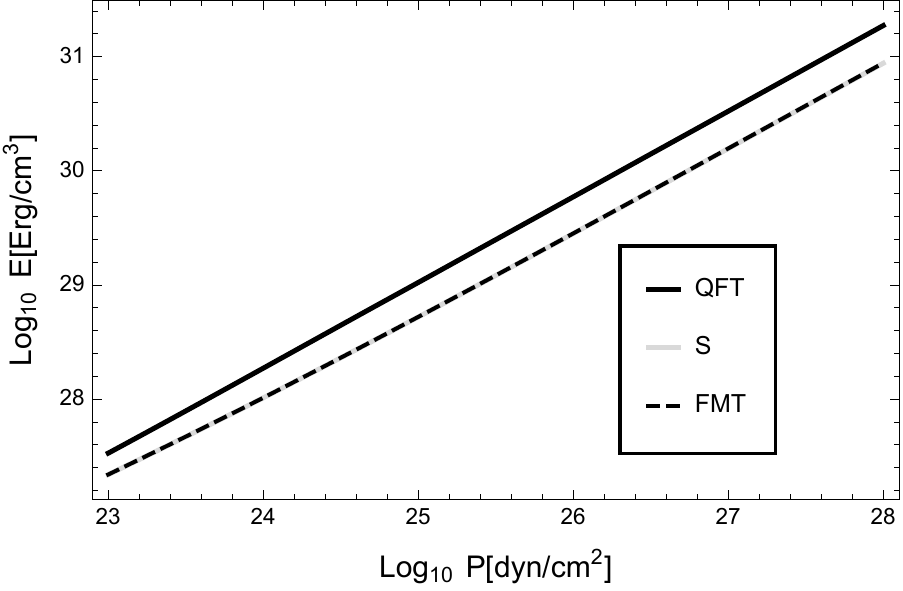} \caption{Comparison of $E(P)$ for carbon ${\rm C}^{12}_{6}$ between the proposed model by Salpeter (S), Feynman-Metropolis-Teller (FMT) and our model using quantum field theory (QFT).}
\label{densvsenergia}
\end{figure}

In Fig.~\ref{densvspressao} we present $P=P(\rho)$, whereas in Fig.~\ref{densvsenergia} we present $E=E(P)$.
In these plots we compare our calculations, labelled as QFT due to the quantum field theory approach, with the results from Chandrasekhar (Ch), Salpeter (S) and Feynman-Metropolis-Teller (FMT).
We also present some relevant data.
In Table.~\ref{table_pressure}, we fix some numerical values for the mass density and present the values of the macro variables for the aforementioned models. 
The data from Salpeter and Feynman-metropolis-Teller cases for the plots and tables are given in Ref~\cite{Rotondo:2011zz}.
Finally, in Table.~\ref{linear_adjust},
we present the linear regression for the EoS $P=P(\rho)$ in the logarithm scale for different chemical elements.
In Table.~\ref{comp_model_log}, 
we present the linear regression for the EoS $P=P(\rho)$ in the logarithm scale for $C^{12}_6$ for the different models we have considered.
\begin{table}[htb!]
\centering 
\begin{tabular}{c c c c c}
\hline\hline
$\rho$ \  & \ $P_{\rm S}$ \ & \ $P_{\rm FMT}$ \ & \ $P_{\rm QFT}$  \\ [0.5ex]
\hline
$10^8$ \ & \ $2.114\times10^{25}$ \ & \ $2.109\times10^{25}$ \ & \ $8.217\times10^{24}$ \\
$10^9$  \ & \ $4.782\times10^{26}$ \ & \ $4.767\times10^{26}$ \ & \ $1.757\times10^{26}$ \\
$10^{10}$ \ & \ $1.042\times10^{28}$ \ & \ $1.037\times10^{28}$ \ & \ $3.783\times10^{27}$ \\
[1ex]
\hline
\end{tabular}
\caption{
Equation of state $P=P(\rho)$ for Carbon for Salpeter (S), Feynman-metropolis-Teller (FMT) and our present model (QFT). 
The pressure $P$ and the mass density $\rho$ are given is units of $\rm g/cm^3$ and $\rm dyn/cm^2$, respectively.}
\label{table_pressure}
\end{table}
\begin{table}[htb!]
\centering 
\begin{tabular}{c c c }
\hline\hline
Element \  & \ Angular coefficient \ & \ Linear coefficient  \\ [0.5ex]
\hline
${\rm He}^4_2$ \ & \ $1.3326$ \ & \ $14.2551$ \\
${\rm C}^{12}_{6}$  \ & \ $1.3333$  \ & \ $14.2446$  \\
${\rm O}^{16}_{8}$ \ & \ $1.3337$ \ & \ $14.2415$ \\
[1ex]
\hline
\end{tabular}
\caption{
 Coefficients of the linear regression for the EoS $P=P(\rho)$ for some elements in logarithm scale. Pressure $P$ and mass density $\rho$ are given in units ${\rm dyn/cm^2}$ and ${\rm g/cm^3}$, respectively.}
\label{linear_adjust}
\end{table}
\begin{table}[htb!]
\centering 
\begin{tabular}{c c c }
\hline\hline
Model \ \  & \ \ Angular coefficient \ \ & \ \ Linear coefficient  \\ [0.5ex]
\hline
Ch \ \ & \ \ 1.56 \ \ & \ \ 12.79 \\
S  \ \ & \ \ 1.60  \ \ & \ \ 12.47  \\
FMT \ \ & \ \ 1.57 \ \ & \ \ 12.71 \\
QFT \ \ & \ \ 1.33 \ \ & \ \ 14.24 \\
[1ex]
\hline
\end{tabular}
\caption{
Coefficients of the linear regression for the EoS $P=P(\rho)$ for ${\rm C}^{12}_6$ in logarithm scale
for  different models: Chandrasekhar (Ch), Salpeter (S), Feynman-Metropolis-Teller (FMT) and the present model using quantum field theory (QFT).
Pressure $P$ and mass density $\rho$ are given in units ${\rm dyn/cm^2}$ and ${\rm g/cm^3}$, respectively.}
\label{comp_model_log}
\end{table}
The data of Table.~\ref{linear_adjust}
show the polytropic feature with respect to the element composition of the lattice for low charge elements,
whereas Table.~\ref{comp_model_log} show a significant difference between our model and the others we have considered.

\subsection{Electron capture limit }
 
In order to point out the applicability limits of our results, we determine the threshold for capture processes~\cite{Yuan:2005tj}, which may alter the ions of the lattice. 
In the case of neutron stars, e.g., it separates the region of the inner crust from the outer crust. 
The calculations are performed using the data from Figs~\ref{densvspressao} and \ref{densvsenergia}.

We can define this threshold from electron capture. 
This phenomenon can be represented by:
\begin{eqnarray}
\epsilon^{\beta}_Z =
\lim_{r\rightarrow R_0} \epsilon_e(\bar{\mu}_e(\vec{r})),
\end{eqnarray}
where $\epsilon_e$ is the plasma energy density,
$R_0$ is the nuclear radius and the limit indicates the energy density in the vicinity of the nucleus (bare nuclei), given by $\epsilon_Z^\beta$.  
In order to compare with other models, we consider the data from Ref.~\cite{Rotondo:2011zz}, which are given in Table~\ref{tabbeta}.
These results indicate no relevant effect in the critical mass density due to the presence of the lattice, which remains with the same order of magnitude.
\begin{table}[htb!]
 \centering 
\begin{tabular}{l l l l l}
\hline\hline
Decay \ & \ $\epsilon^{\beta}_{Z}$ \ \ & \  $\rho_{crit}^{\beta ,\rm unif}$ \ \ & \ \ $\rho^{\beta,\rm rel.\,FMT}_{\rm crit}$\ \ & \ \ $\rho^{\beta, \rm QFT}_{\rm crit}$ \\ 
\hline
${\rm He}^{4}\rightarrow {\rm H}^3+{\rm n}\rightarrow 4{\rm n}$  \ & \ $20.596$ \  &  \  $1.37\times 10^{11}$ \ & \ $1.39\times 10^{11}$ \ \ & \ $1.26\times 10^{11}$ \\
${\rm C}^{12}\rightarrow {\rm B}^{12}\rightarrow {\rm Be}^{12}$  \ & \ $13.370$  \  &  \  $3.88\times 10^{10}$\ & \ $3.97\times 10^{10}$\ & \ $2.49\times 10^{10}$ \\
${\rm O}^{16}\rightarrow {\rm N}^{16}\rightarrow {\rm C}^{16}$  \ & \ $10.419$  \  &  \  $1.89\times 10^{10}$ \ & \ $1.94\times 10^{10}$ \ & \ $8.48\times 10^9$ \\
[1ex]
\hline
\end{tabular}
\caption{Comparison between values presented in Ref.~\cite{Rotondo:2011zz} with the model presented in this work. The density being $\rho$ written in ${\rm g/cm^3}$ and $\epsilon^{\beta}_Z$ in MeV.}
\label{tabbeta}
\end{table}

\section{Conclusions}
\label{sec7}
 
In this work we investigate the effects of a body-centered cubic crystal lattice of ions in the QED plasma, which is important to accurately describe the structure of compact objects such as white dwarf stars and the crust of neutron stars.
Our main goal is to obtain how the EoS of the plasma is affected by the presence of the crystal lattice. 
Our calculation scheme is to apply linear response theory in the context of quantum field theory at finite temperature, where the QED plasma responds to the external influence of the classical crystal lattice. 
For simplicity, we assume that the crystal lattice remains static.
This approach is more robust than the usual ones considered in the literature, which becomes clear when we show the importance of computing the influence of the electric potential of arbitrary distant ions of the lattice and not only of the nearest neighbors.
This is detailed  in   Appendix~\ref{appendixrcellsec}, where the influence of electric potential of more distant ions proves to be important in obtaining the Wigner-Seitz radius of the lattice cells.
Additionally, it provides a tools to investigate the formation of the lattice and is more precise than the Coulomb parameter of Eq.~\eqref{cristalizacao}, frequently used in the literature.

Changes in chemical potential of the electrons induced by the crystal lattice interaction with QED plasma leads to an non homogeneous equilibrium distribution of electrons inside WS cell (see local hydrostatic equilibrium condition established in Appendix~\ref{equilibriumappendix}).

At the hydrodynamics scale,  we have introduced macro thermodynamics variables associated to the cell as a whole. These macro variables are relevant to obtain the EoS the medium for future applications.

The results are presented by some representative situations of compact stars, the  relativistic degenerate gas. Results for the lattice influence  inside each cell and are given in the figures of Sec.~\ref{results_interior}.
The EoS calculation  are given Sec.~\ref{results_exterior} in some figures and tables, which show some differences with respect to other models of the literature.

The results of this work can be widely applied in astrophysical models of neutron stars and white dwarfs in order to provide corrections to stellar structure. 
It can be used as a basis to solve Tolman-Oppenheimer-Volkoff (TOV) equations and some modifications can be made to include particle emissivity, magnetic fields, MCP composition, nuclear processes, crystal lattice vibration, different lattice structures like fcc and others. 

Finally, as an important caveat we mention that linear response theory is a next-to-leading order approximation, which limits our model to weak electric field generated by the ions and, consequently, the model needs to be carefully inspected before future applications to fully ionized ions with a larger $Z$ number.

In the case of neutron stars, the process of neutrino emissivity is a very relevant process for the cooling of the star. It would be interesting to consider the effect of the crystal lattice, i.e., the modified EoS in the emissivity rate. 
This will be done in a work in progress.

\section*{Acknowledgement}
This work is a part of the
project INCT-FNA proc.~No.~464898/2014-5.
S.~V.~C.~Ramalho is supported by Coordena\c{c}\~ao de Aperfei\c{c}oamento de Pessoal de N\'{\i}vel Superior (CAPES).
S.~B.~Duarte is   
partially supported by Conselho Nacional de Desenvolvimento
Cient\'{\i}fico e Tecnol\'ogico (CNPq).

\appendix

\section{Determination of the Wigner-Seitz cell radius} \label{appendixrcellsec}

In the literature, the crystal lattice formation is described using the Coulomb parameter, which was introduced in Sec.~\ref{sec3}.
In this appendix, we consider an alternative determination of the WS radius, $R_{\rm cell}$, when we have the ion lattice interacting with the non homogeneous electron plasma. 
The interaction is described by linear response theory in the context of quantum field theory.

Here we have computed $R_{\rm cell}$, considering the interaction between distant neighbors and contrasted to the results presented in the literature, which take into account only the interaction between first neighbours.

The charge neutrality of the WS cell filled by a non homogeneous electron plasma distribution, $n_e$, is given by
\begin{eqnarray}
\int_{V_{\rm WS}} n_e\,(\bar{\mu}_e,T_e,Z)\,dV=Z,
\label{intndvws}
\end{eqnarray}
where $V_{\rm WS}$ is the volume of a WS cell and $\bar{\mu}_e=\bar{\mu}_e(\vec{r})$.
Using spherical coordinates, we can write Eq.~\eqref{intndvws} as:
\begin{eqnarray}
\int_0^{2\pi}
\int_{0}^{\pi}
\int_{R_0}^{R_{\rm cell}}
n_e(\bar{\mu}_e, T_e,Z_{\rm eff})
\,r^2\sin\theta\, dr\,d\theta\, d\phi=Z_{\rm eff},
 \label{intdensidade}
\end{eqnarray}
where $R_0 \sim A^{1/3}$ is the nuclear radius.
Defining the result of the LHS of the latter equation as $f(\mu_e, T_e,Z_{\rm eff},R_{\rm cell})$, we can write define $g(\mu_e,T_e,R_{\rm cell},Z_{\rm eff})$ as follows:
\begin{eqnarray}
g(\mu_e,T_e,R_{\rm cell},Z_{\rm eff})=f(\mu_e,T_e,R_{\rm cell},Z_{\rm eff})-Z_{\rm eff}=0. 
 \label{gz}
\end{eqnarray}
For each value of $\mu_e$, $T_e$ and $Z$, we obtain a corresponding value of $R_{\rm cell}$. Fixing $\mu_e$, $T_e$ and $Z$, we solve the numerical integral for a range of values of $R_{\rm cell}$ and plot the function given in Eq.~\eqref{gz} in order to determine its roots.

\begin{figure}[htb!]
\centering 
\includegraphics[width=10cm]{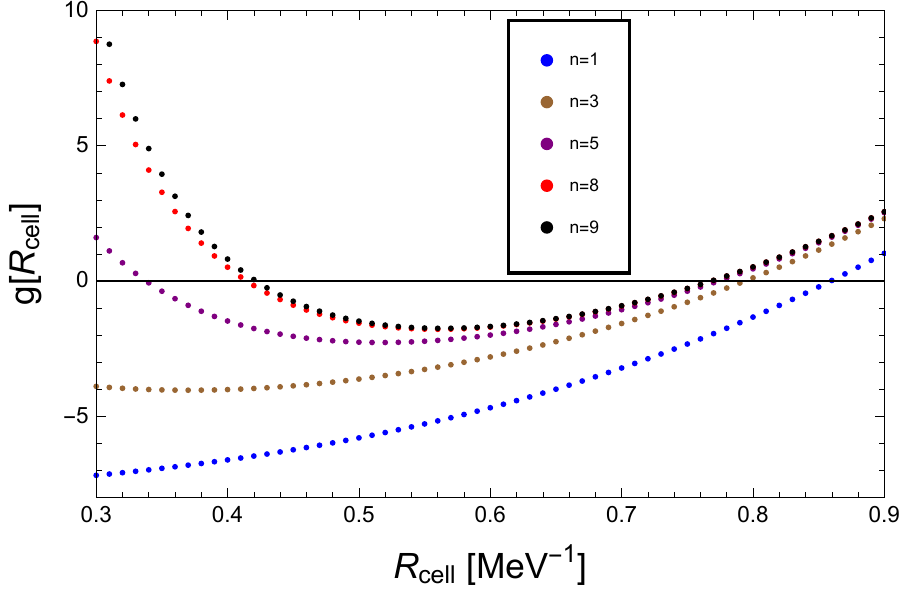} \caption{Numerical plot for Eq.~\eqref{gz} for the degenerate relativistic gas limit. We chose the representative values $Z_1=Z_2=8$, $A=16$, $\mu_e= 6 \, {\rm MeV}$ and for neighborhood orders:  $n=1$ (blue), $n=3$ (brown), $n=5$ (purple), $n=8$ (red) and $n=9$ (black).}
\label{highdensn5}
\end{figure}

Particularly, we now analyse the relativistic degenerate gas limit from Sec.~\ref{sec5} by solving Eq.~\eqref{gz} numerically for different values of $\mu_e$ and $Z_{\rm eff}$ and different neighborhood orders. From the results of Fig.~\ref{highdensn5}, we notice that the neighborhood order affects the value of $R_{\rm cell}$.
Noticing that the root of the function $g$ converges to a specific value as we increase the neighborhood order, we obtain $R_{\rm cell}\simeq 0.7683\, {\rm MeV^{-1}}= 151.35\, {\rm fm}$. 
It is important to mention that the smaller root of Fig.~\ref{highdensn5} has no physical meaning and is ignored.

\section{Local hydrostatic equilibrium}\label{equilibriumappendix}

The interaction between the lattice ions and electron in the plasma imposes a change in the chemical potential of the electrons as describe in Ref.~\cite{han}. An effective chemical potential can be defined,
\begin{eqnarray}
 \bar{\mu}_e(r)=\mu_e+\delta\mu_e(r), \label{A0}
\end{eqnarray}
where $\delta\mu_e(r)$ denotes a chemical potential perturbation imposed by the presence of the lattice.

This calculation of $\delta\mu_e(r)$ can be achieved from the local hydrostatic equilibrium condition, which requires the balance of forces according to the following equation:
\begin{eqnarray} \label{equihydro}
\vec{\nabla} P_e\left(\bar{\mu}_e(r),T\right)=-e \ n_e(\bar{\mu}_e(r),T) \vec{E}_{{\rm net}}(\vec{r}),
\end{eqnarray}
where $P_e\left(\bar{\mu}_e(r),T\right)$ and $n_e(\bar{\mu}_e(r),T)$ are the pressure and number density, respectively, of the non homogeneous plasma and $\vec{E}_{{\rm net}}(\vec{r})$ is the screened electric field. 
Writing the electric field in terms of the potentials $\Phi$ and $\vec{A}$, $\vec{E}=-\vec{\nabla} \Phi-\partial_t \vec{A}$, and substituting it in Eq.~\eqref{equihydro}, one gets:
\begin{eqnarray}\label{equihydro2}
\left(\frac{\partial P_e \left(\bar{\mu}_e(r),T\right)}{\partial\bar{\mu}_e(r)}\right)\vec{\nabla}\delta\mu_e(r)=e \ n_e(\bar{\mu}_e,T)\left(\vec{\nabla} \Phi(\vec{r})+\partial_t \vec{A}(\vec{r}) \right),
\end{eqnarray}
which results:
\begin{eqnarray}
\label{nabladeltamu}
-\nabla\delta\mu_e(r)
+
e
\left(\vec{\nabla} \Phi(\vec{r})+\ \partial_t \vec{A}(\vec{r})\right)=0.
\end{eqnarray}
In this work, we only consider the effects of the static electric field of the lattice ions.
Therefore, we can eliminate the third term from Eq.~\eqref{nabladeltamu}, leading to the equilibrium condition of the plasma,
\begin{eqnarray}\label{deltamueAr}
\delta\mu_e(r)= e\Phi(r)+{\rm constant}.
\end{eqnarray}


\begin{thebibliography}{99}



\bibitem{Glendenning} 
N.~K.~Glendenning, {\it Compact Stars, Nuclear Physics, Particle Physics, and General Relativity}, Springer-Verlag, New York (1996).

\bibitem{Camenzind}
M.~Camenzind, {\it Compact Objects in Astrophysics: White Dwarfs, Neutron Stars and Black Holes}, Springer, Berlin, Heidelberg (2007).


\bibitem{Chamel:2008ca}
N.~Chamel and P.~Haensel,
Physics of Neutron Star Crusts,''
Living Rev. Rel. \textbf{11}, 10 (2008).


\bibitem{Lattimer:2000nx}
J.~M.~Lattimer and M.~Prakash,
``Neutron star structure and the equation of state,''
Astrophys. J. \textbf{550}, 426 (2001).


\bibitem{measuriment}
A.~L.~Kritcher, D.~C.~Swift, T.~D{\"o}ppner, et al., ``A measurement of the equation of state of carbon envelopes of white dwarfs'', Nature {\bf 584}, 51-54 (2020). 


\bibitem{Ozel:2016oaf}
F.~\"Ozel and P.~Freire,
``Masses, Radii, and the Equation of State of Neutron Stars,''
Ann. Rev. Astron. Astrophys. \textbf{54}, 401-440 (2016).


\bibitem{Haensel:2007yy}
P.~Haensel, A.~Y.~Potekhin and D.~G.~Yakovlev,
``Neutron stars 1: Equation of state and structure'',
Astrophys. Space Sci. Libr. \textbf{326}, pp.1-619 (2007).


\bibitem{baiko} 
D.~A.~Baiko, 
``Coulomb crystals in neutron star crust'',
Journal of Physics: Conference Series~{\bf 496}, 012010 (2014).


\bibitem{nature}
H.~M.~Van~Horn, ''The crystallization of white dwarf stars''. Nat Astron 3, 129-130 (2019).


\bibitem{bellac} 
M.~Le~Bellac, {\it Thermal Field Theory}, Cambridge University Press, New York (2011).

\bibitem{kapusta}
J.~I.~Kapusta and C.~Gale,
{\it Finite-Temperature-Field Theory Principles and Applications}, Cambridge University Press, New York (2006).


\bibitem{Rotondo:2011zz}
M.~Rotondo, J.~A.~Rueda, R.~Ruffini and S.~S.~Xue,
``The Relativistic Feynman-Metropolis-Teller theory for white dwarfs in general relativity,''
Phys. Rev. D \textbf{84}, 084007 (2011).



\bibitem{Baldo:2009we}
M.~Baldo and C.~Ducoin,
``Plasmon excitations in homogeneous neutron star matter,''
Phys. Atom. Nucl. \textbf{72}, 1188-1196 (2009).

\bibitem{Braaten:1993jw}
E.~Braaten and D.~Segel,
``Neutrino energy loss from the plasma process at all temperatures and densities,''
Phys. Rev. D \textbf{48}, 1478-1491 (1993).


\bibitem{Yakovlev:2000jp}
D.~G.~Yakovlev, A.~D.~Kaminker, O.~Y.~Gnedin and P.~Haensel,
``Neutrino emission from neutron stars'',
Phys. Rept. \textbf{354}, 1 (2001).

\bibitem{Altherr:1993hb}
T.~Altherr and P.~Salati,
``The Electric charge of neutrinos and plasmon decay,''
Nucl. Phys. B \textbf{421}, 662-682 (1994).

\bibitem{Chandrasekhar:1931ih}
S.~Chandrasekhar,
``The maximum mass of ideal white dwarfs,''
Astrophys. J. \textbf{74}, 81-82 (1931).


\bibitem{Salpeter:1961zz}
E.~E.~Salpeter,
``Energy and Pressure of a Zero-Temperature Plasma,''
Astrophys. J. \textbf{134}, 669-682 (1961).


\bibitem{Hossain:2019teg}
G.~M.~Hossain and S.~Mandal,
``Revisiting equation of state for white dwarfs within finite temperature quantum field theory,''
[arXiv:1904.09174 [astro-ph.HE]].



\bibitem{Drewsen} 
M.~Drewsen, 
``Ion Coulomb crystals'', Physica B: Condensed Matter, {\bf 460}, 105-113 (2015).




\bibitem{quantumcrystal} 
M.~Bonitz, P.~Ludwig, H.~Baumgartner, et al.,
``Classical and quantum Coulomb Crystal'',
Physics of Plasmas {\bf 15}, 055704 (2008). 


\bibitem{simulations}
M.~D~Jones, D.~M.~Ceperley,
``Crystallization of the One-Component Plasma at Finite Temperature,''
 Phys. Rev. Lett. \textbf{76(24)}, 4572-4575 (1996).





\bibitem{Ruester:2005fm}
S.~B.~Ruester, M.~Hempel and J.~Schaffner-Bielich,
``The outer crust of non-accreting cold neutron stars,''
Phys. Rev. C \textbf{73}, 035804 (2006).

\bibitem{walecka}
A.~L.~Fetter and J.~D.~Walecka, {\it Quantum Theory of Many-Particle Systems}, McGraw-Hill, New York (1971).



\bibitem{stella}
R.~Ruffini, L.~Stella, ''Some comments on the relativistic Thomas-Fermi model and the Vallarta-Rosen equation,'' Physics Letters B, \textbf{102}(6), 442-444 (1981).

\bibitem{flambaum}
V. V. Flambaum and J. S. M. Ginges , "Electric field distribution in nuclei produced by the P,T-odd nuclear schiff moment", AIP Conference Proceedings \textbf{596}, 232-245 (2001).

\bibitem{Kapusta:1988fi}
J.~I.~Kapusta and T.~Toimela,
``Friedel Oscillations in Relativistic {QED} and {QCD},''
Phys. Rev. D \textbf{37}, 3731 (1988).


\bibitem{han}
G.~Han and H.~Wang, 
``Generalized expression of chemical potential with influence of external fields and its applications: Effect of charged particles on droplet condensation'',
Fluid Phase Equilibria {\bf 338}, 269 (2013).


\bibitem{chung}
K.~C.~Chung, {\it Introdu\c{c}\~ao \`a F\'{\i}sica Nuclear}, EDUERJ, Rio de Janeiro (2001).


\bibitem{Tolman:1939jz}
R.~C.~Tolman,
``Static solutions of Einstein's field equations for spheres of fluid,''
Phys. Rev. \textbf{55}, 364-373 (1939).


\bibitem{Oppenheimer:1939ne}
J.~R.~Oppenheimer and G.~M.~Volkoff,
``On Massive neutron cores,''
Phys. Rev. \textbf{55}, 374-381 (1939).






\bibitem{Yuan:2005tj}
Y.~F.~Yuan,
``Electron positron capture rates and the steady state equilibrium condition for electron-positron plasma with nucleons,''
Phys. Rev. D \textbf{72}, 013007 (2005).






\end{thebibliography}
\end{document}